\documentstyle[aps,twocolumn,prl,epsf,rotating,amssymb]{revtex} %with lines(3) of '%+++++'
\begin{document}

\draft
\twocolumn[    %+++++
\hsize\textwidth\columnwidth\hsize\csname @twocolumnfalse\endcsname    %+++++

\title{Stripe phases in the two-dimensional Falicov-Kimball model}
\author{R. Lema\' nski$^1$,
J.K. Freericks$^2$ and G. Banach$^{1,3}$}
\address{$^1$Institute of Low Temperature and Structure Research,
Polish Academy of Sciences, Wroc\l aw, Poland \\
$^2$Department of Physics, Georgetown University, Washington, DC
20057, U.S.A. \\
Daresbury Laboratory, Cheshire WA4 4AD, UK}

\date{\today}
\maketitle

\widetext
\begin{abstract}
The observation of charge stripe order in the doped nickelate and cuprate
materials has motivated much theoretical effort to understand the underlying
mechanism of the stripe phase.  Numerical studies of the Hubbard model
show two possibilities: (i) stripe order arises from a tendency toward
phase separation and its competition with
the long-range Coulomb interaction or (ii) stripe
order inherently arises as a compromise between itinerancy and magnetic
interactions. Here we determine the restricted phase diagrams of the
two-dimensional Falicov-Kimball model and see that it displays rich behavior
illustrating both possibilities in different regions of the phase diagram.
\end{abstract}

]      %+++++

\narrowtext
%============================================================================
The discovery of charge stripes in the nickelate~\cite{nickel} and
cuprate~\cite{cuprate} materials has encouraged much theoretical work
to explain the underlying physical principles behind the stripe order
and to examine whether the stripes are related to the mechanism for
high-temperature
superconductivity.  There are two schools of thought for the physics that
drives stripe formation: (i) Kivelson, Emery, and coworkers~\cite{kivelson}
propose that
strongly correlated systems have a natural tendency toward phase separation
and the inhomogeneous
spatial charge ordering arises from a competition between this
tendency to phase separate and the long-range Coulomb interaction which
does not allow the electron density to stray too far from its average;
and (ii) Scalapino and White~\cite{scalapino} propose that the stripe order
arises from a competition between kinetic and exchange energies in a doped
antiferromagnet which does not require long-range Coulomb forces to stabilize
the stripes.  Despite a large amount of numerical work ranging
from high-temperature expansions~\cite{hight}, to Monte Carlo
simulations~\cite{qmc}, to exact diagonalization~\cite{diag},
as well as semiclassical Hartree-Fock theory~\cite{HF},
no consensus has been reached about the region of stability for the phase
separated states or the mechanism for stripe formation.

Here we take an alternate point of view.  Rather than try to prove phase
separation in the Hubbard or $t-J$ models, we choose an even simpler
model---the spinless Falicov-Kimball model~\cite{falicov_kimball}, which
can be analyzed exactly.  The relation of the Falicov-Kimball model to the
Hubbard model is analogous to the relation between the Ising and the
Heisenberg models of magnetism (the Falicov-Kimball model can be viewed
as a Hubbard model where the down-spin electrons are frozen and do not
hop).  The Hamiltonian is
\begin{equation}
H=-t\sum_{\langle x,y\rangle}c^{\dagger}_{x}c_{y}+
U\sum_{x} c^{\dagger}_{x}c_{x}w_{x},
\end{equation}
with $c^{\dagger}_{x}$ ($c_{x}$) the creation (annihilation) operator
for a
spinless electron at site $x$ and $w_{x}$ is a classical variable that
denotes
the presence (absence) of an ion at site $x$ when it is equal to 1 (0),
respectively. The hopping occurs between nearest-neigbors on a square
lattice and the interaction strength is denoted by $U$.
For any given configuration of ions $\{w_{x}\}$ the ground state for $N_e$
electrons is determined
by diagonalizing a one-body operator given by the above Hamiltonian, and filling
in the lowest $N_e$ states.  We typically are interested
in the ground-state configuration of the ions for a given number of electrons
$N_e=\sum_{x}\langle c^{\dagger}_{x}c_{x}\rangle$ and a given
number of ions $N_i=\sum_{x}w_{x}$ (or their densities $\rho_e=N_e/N$
and $\rho_i=N_i/N$ respectively).

The possibility of phase separation when $U\rightarrow\infty$ was proposed
in 1990~\cite{freericks_falicov} and is called the segregation principle.
It was proved in the one-dimensional case~\cite{lemberger},
in the infinite-dimensional case~\cite{freericks_infd}, and recently in the
general case~\cite{freericks_lieb_ueltschi}.  This phase separation is
a special type of phase separation, often referred to as the segregated phase,
where the electrons and ions avoid each other, and reside in separate domains.
It is the analog of the ferromagnetic state in the Hubbard model.
At half filling for the electrons and the ions, the ground state is known to
be the chessboard phase~\cite{lieb_kennedy,brandt_schmidt}.  This state is
the analog of the antiferromagnetic state in the Hubbard model.

The question we pose is, if we fix the number of electrons to
equal the number of ions, then what are the stable phases as a function
of the number of electrons.  This problem is the analog of doping the
$S_z=0$ phase of the Hubbard model away from half filling.  Since we know
the system phase separates as $\rho_e\rightarrow 0$ and is in the chessboard
phase for $\rho_e\rightarrow 1/2$, we have the interesting situation of
determining how the transition is made from the segregated phase to the
chessboard phase, which are two phases that are about as different from
each other as possible.

It is well known that the ground-state phase diagram of the Falicov-Kimball
model typically includes a large number of different phases as functions
of the particle concentrations and the interaction strength.  As such, it
seems unlikely that one can rigorously analyze ground-state phases in the
general case, although a number of interesting results exist for special
cases~\cite{2d_known} (usually for large $U$).  Instead, we work with the
restricted phase diagram technique, where we consider all possible periodic
phases for which the number of sites per unit cell $N_0$ is less than
or equal to $N_c=16$. The technique first identifies all nonequivalent
periodic phases, which number 23 755 in our case. Then for each periodic
phase in our trial set, we calculate the total energy at a given value of $U$.
Since, unlike the 1D case (see \cite{RL}), there are no exact formulae known
for the density of states of general periodic phases in 2D, we have performed
a numerical solution of the corresponding eigenvalue problem to determine the
bandstructure (see Ref.
\cite{watson_lemanski} for the details). This involves finding the eigenvalues
of an $N_0-dimensional$ matrix for each value of $k$ in a two-dimensional
grid covering the Brillouin zone. Our calculations were performed with
a $k-space$ grid of 110$\times$110 points for each phase. Then the grand canonical
phase diagram is constructed as a function of the electron and ion chemical
potentials, and finally the grand-canonical phase diagram is translated
to the canonical phase diagram for arbitrary $\rho_i$ and $\rho_e$.
This procedure assures thermodynamical stability of all phases (both periodic
and their mixtures) present in the resulting canonical phase diagram.
The stability problem is discussed extensively in Ref. \cite{GJL}, where the
canonical phase diagrams of the 1D Falicov-Kimball model were studied.
Finally, we make the restriction $\rho_i=\rho_e$ in the canonical phase diagram
to analyze the problem at hand.

The ground-state phase diagram is quite complex.  We find many different
stable phases occur, which can be classified into a number of different
categories: (i) {\em the empty lattice} ($\rho_i=0$ and $\rho_e\ne 0$) denoted {\bf E};
(ii) {\em the full lattice} ($\rho_i=1$ and $\rho_e=0$) denoted {\bf F};
(iii) {\em the chessboard phase} ($\rho_i=\rho_e=1/2$ and ions occupy
the A sublattice only) denoted {\bf Ch};
(iv) {\em diagonal neutral} ($\rho_i=1-\rho_e$) {\em stripe phase}s (the ions are
arranged as diagonal chessboard phases separated by fully occupied striped
regions with a slope of one, or equivalently, Ch phases separated by diagonal
antiphase boundaries) denoted {\bf DNS};
(v) {\em diagonal non-neutral} ($\rho_i\neq 1-\rho_e$) {\em stripe phases}
(the ions are arranged as diagonal chessboard phases but separated by empty striped
regions with a slope of one) denoted {\bf DS};
(vi) {\em axial non-neutral stripes} ($\rho_i\neq 1-\rho_e$ and ions arranged in
stripes parallel to the y-axis and translationally invariant along the
axis) denoted {\bf AS}; (vii) {\em axial non-neutral
chessboard stripes} ($\rho_i\neq 1-\rho_e$ and ions arranged in stripes of
the chessboard phase oriented parallel to the y-axis) denoted {\bf AChS}; (viii)
{\em other neutral phase}s ($\rho_i=1-\rho_e$ but the arrangement is not in
any simple stripe-like phase) denoted {\bf N}; (ix) non-neutral
{\em four-molecule phases} (where $\rho_i\neq 1-\rho_e$ and
empty sites are arranged out of ``bound'' four-molecule squares) denoted {\bf 4M};
and (x) truly two-dimensional non-neutral arrangements of ions
(where $\rho_i\neq 1-\rho_e$ and the ions are
arranged in a fashion that is not stripe-like, but rather requires a
two-dimensional unit cell to describe it) denoted {\bf 2D}. Generically, the
phase diagram includes mixtures of two or three of the periodic phases (iii-x),
or of one or two periodic phases and the empty lattice (i).  The empty lattice is
usually needed in the phase mixtures to ensure that the average electron
and ion fillings are equal for the mixtures.  For occasional values of
the filling, we do find single periodic phases to be stable, but this
feature is rare.  When the filling is sufficiently far from half
filling, the system is in the segregated phase, which is a mixture of
the empty lattice ($\rho_i=0$, $\rho_e\ne 0$) and of the full lattice
($\rho_i=1$, $\rho_e=0$).

The number of phases  stable or appearing in mixtures
tends to grow as $U$ is decreased in magnitude.
For $U=8$ there are 25 phases: E (1);  F(1); Ch (1); DNS (6); and N (16).
For $U=6$ there are 30 phases: E (1); F (1); Ch (1); AS (20); N (6); and 2D (1).
For $U=4$ there are 42 phases: E (1); F (1); Ch (1); AS (35); N (1); and 2D (3).
For $U=2$ there are 38 phases: E (1); F (1); Ch (1); AS (14); DNS (4); 4M (2);
and 2D (15).
For $U=1$ there are 50 phases: E (1); F (1); Ch (1); DS (1); AS (9); AChS (6);
4M (3); and 2D (28).

\begin{figure}[htb]

\epsfxsize=8.0cm \epsfbox{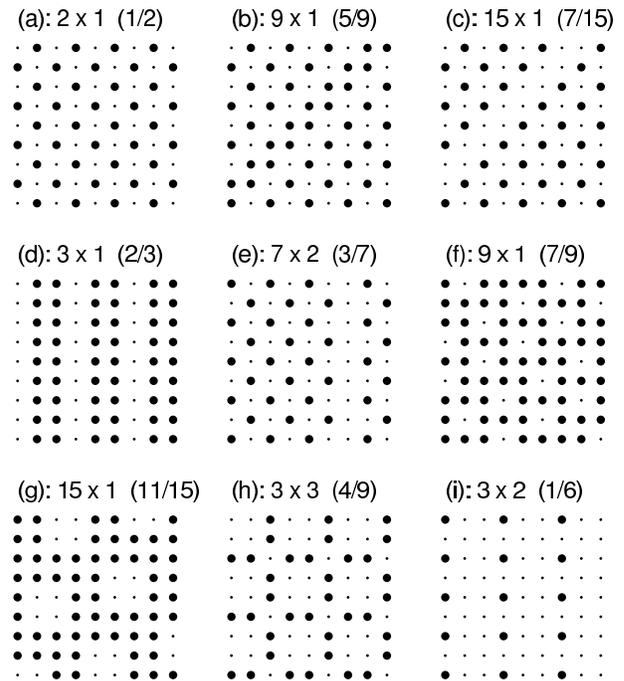}

\caption{\it Characteristic configurations representing phases displayed
in the phase diagram: (a) Ch, (b) DNS, (c) DS, (d) AS, (e) AChS, (f) N, (g) 4M,
(h) 2D, and (i) another 2D.
The large dots correspond to  ion-occupied sites and the small dots
correspond to ion-vacant sites.
The numbers written above each  configuration indicate the dimensions
of the unit cell; the corresponding ion densities are given in parentheses.}
\label{fig1}
\end{figure}

The empty lattice, the full lattice and the
chessboard phase are present in the phase diagram
for all $U$; the diagonal neutral stripes
generally for large $U$ ($U> 7$) (occasionally they can appear for moderate $U$
as well---as occurs for $U=2$);
the axial stripes for moderate and small $U$ ($U< 7$); the axial
chessboard stripes for small $U$ $(U<3$); the other neutral phases
for large $U$ ($U>4$); the four-molecule phases for small $U$ ($U<3$);
and the two-dimensional phases for moderate and small $U$ ($U<7$)
(growing significantly in number as $U$ decreases).  The total number  of phases
appearing in the phase diagram are too large to illustrate here.  Instead,
we choose a number of illustrative examples of each of the different types.

The empty lattice ({\bf E}) corresponds to a lattice with no ions and the full
lattice ({\bf F}) corresponds to a lattice fully occupied by ions. The chessboard
phase ({\bf Ch}) corresponds to the case where ions (large dots) occupy
the A sublattice and empty sites (small dots) occupy the B sublattice as shown
in Fig.~1(a). A diagonal neutral stripe phase ({\bf DNS}) is shown in Fig.~1(b).
It is typical of the phases we see with regions of chessboard phase separated
by antiphase boundaries.  This configuration has $\rho_i=5/9$ and $\rho_e=4/9$
and every ninth diagonal is an antiphase boundary. An example of a nonneutral
diagonal stripe phase ({\bf DS}) which has $\rho_i=7/15$ and $\rho_e=7/15$ is shown
in Fig.~1(c). Like the neutral stripes, these have antiferromagnetic regions
but they are separated by diagonal empty (and not full ) lattice
stripes; the empty lattice stripes are width two now.
An axial stripe phase ({\bf AS}) is shown in Fig.~1(d).
It is typical of these phases with fully occupied (ferromagentic) stripes
separated by the empty lattice. This configuration has $\rho_i=2/3$
and $0.189 < \rho_e < 0.283$; other stable phases have wider bands of occupied
stripes, which are always separated by an empty lattice stripe of width one.
An example of an axial chessboard stripe phase ({\bf AChS}) is shown in Fig.~1(e).
These phases consist of regions of the chessboard phase separated by width one
empty lattice stripes.  The width of the chessboard phase ranges all the way
down to two.  This particular phase has $\rho_i=3/7$ and $\rho_e=3/7$.
An example of an other neutral phase ({\bf N}) is depicted in Fig.~1(f).
These phases can be viewed as striped phases, alternating occupied (ferromagnetic)
and empty lattice stripes, with the stripes having slope different from 0, 1,
or $\infty$.  In this example, $\rho_i=7/9$ and $\rho_e=2/9$; the slope of the
empty lattice stripe is 1/2. A four-molecule phase ({\bf 4M}) is shown in Fig.~1(g).
These phases are two-dimensional tilings of four-molecule empty-site squares inside
an occupied ``latticework''.  This phase has $\rho_i=11/15$ and $\rho_e=1/15$.
The truly two-dimensional phases ({\bf 2D}) are more difficult to classify---some,
like in Fig.~1(h), have no stripe-like character at all
($\rho_i=4/9$ and $\rho_e=4/9$), others can be described with a stripe-like
picture such as in Fig.~1(i) ($\rho_i=1/6$ and $0.32 < \rho_e < 0.34$) which can be
viewed either as slope 2/3 ferromagnetic stripes in an empty lattice, or as
antiferromagnetic axial stripes, etc.; in the end we classify these phases as
two-dimensional. The summary of all of the stable phases will appear
in a longer publication\cite{lemanski_freericks_long}.

\begin{figure}[htb]
\epsfxsize=7.5cm
\begin{turn}{-90}
\epsfbox{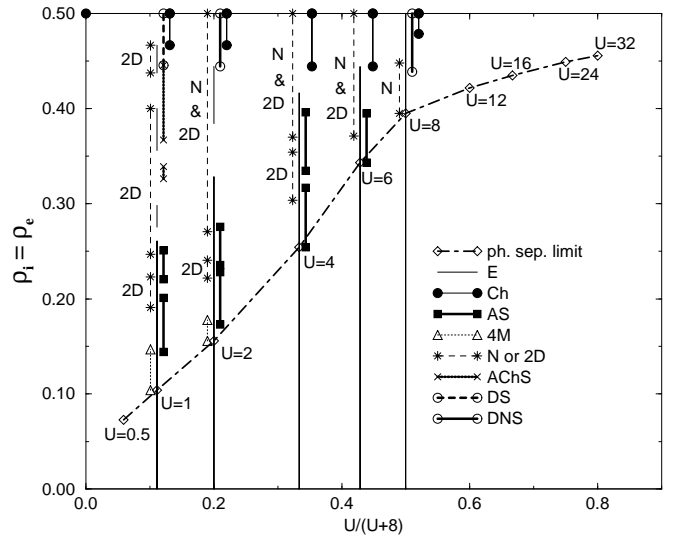}
\end{turn}

\caption{\it Phase diagram of the 2D FKM 
along the $\rho_e = \rho_i$ line for a set of $U$ values.
The dot-dashed line separates the segregated phase and mixtures of the $E$
phase with periodic phases. Below this line only the E and F phases
coexist (in fact the F phase is also stable slightly above the line,
forming three-component mixtures within narrow intervals of densities).
Vertical line segments mark intervals of the
densities where the corresponding phases (of a given class) are involved in the
formation of stable phases. (All the
segments, except for those corresponding to the $E$ phase, are moved
slightly right or left of the value of $U/(U+8)$.)
Bold lines correspond to various types of stripe phases.
If only one category is marked within an interval, only phases belonging
to this category can form stable phases (either single periodic phases or
mixtures of two or three phases from within the same class). If segments
corresponding to different categories
overlap within a given interval, then only mixtures of phases belonging
to those categories can form stable phases.}
\label{fig2}
\end{figure}
The phase diagram is shown in Fig.~2 and its simplified, schematic version,
in Fig.~3.  The phase boundary between the segregated
phase and mixtures with periodic phases, approaches half filling as $U$ gets
large, as expected.  When $U=8$, we find the mixtures, when doped just away from
half filling, occur between the chessboard phase, the empty lattice and diagonal
neutral stripe phases.  This picture is similar to those that propose the phase
separation scenario, but there is no requirement of the long-range Coulomb
interaction to generate the stripes---they also occur as part of the periodic
phases that compose the different stable mixtures.  As the system is doped
further away from half filling, mixtures with other neutral phases occur, before
the system fully phase separates.

As $U$ is reduced, the phase diagram becomes more
complicated.  Near half filling, the chessboard phase is always one of the
phases in the stable mixtures, but we find the empty lattice and diagonal neutral
phases disappear and are replaced by other neutral or 2D phases in the
mixtures.  Then as $U$ is reduced further, the diagonal neutral
stripe phases occasionally re-enter into the mix, replacing the N or 2D phases,
as can be seen for $U=2$.
Finally, for smaller $U$, the mixtures are between the diagonal
non-neutral stripes and the chessboard
phase only.  Farther away from half filling, the behavior is even more complex,
with axial non-neutral
stripes first entering near the segregation boundary, and then the 4M phases
appearing as $U$ is reduced further.  The axial neutral chessboard phases also
appear for small $U$.  It is
in this moderate-to-small $U$ region where we see many stripe phases form
due to a delicate balance between kinetic-energy and potential-energy effects.
This is the alternate scenario for stripe formation, that does not require
phase separation or the long-range Coulomb interaction.

\begin{figure}[htb]
\epsfxsize=7.5cm
\epsfbox{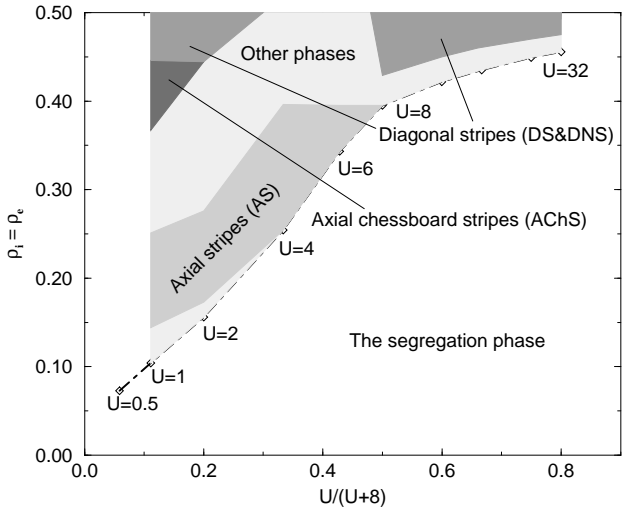}
\caption{\it Schematic phase diagram with highlighted regions of stability of
the stripe phases.}
\end{figure} 

In summary, we have calculated the restricted phase diagram of the two-dimensional
Falicov-Kimball model.  We constrained the system to have the same number
of localized and itinerant particles, which is the analog of the $S_z=0$
state of the Hubbard model.  We find that, in addition to the phase separation
of the segregated phase, the system generically forms a number
of different stripe phases.  For large $U$, we find the stripes to be diagonal
stripes and appearing only close to half filling.  As $U$ is reduced, we find
diagonal stripes changing into axial chessboard stripes and then axial stripes
as the system is doped further away from half filling.  In addition, we find
a number of truly two-dimensional phases present as well.  While we cannot say
anything about what happens in the Hubbard model itself, our results suggest
that one should expect a complex phase diagram when stripe phases are present
and see a competition between the stability of the stripes and other, more
two-dimensional, structures.  Of course, we can only speculate
on the behavior of the Hubbard model, but this exact solution of the
Falicov-Kimball model shows that stripe formation is generically a complicated
occurrence, that can have a number of competing states close in energy. 
We do expect, however, that the phase diagram simplifies
for the Hubbard model, since the system can quantum-mechanically fluctuate between
different low-energy ``ionic'' configurations.

{\it Acknowledgements:}
R.L. and G.B. acknowledge support from the Polish State Committee for Scientific
Research (KBN) under Grant No. 2P03B 131 19 and J.K.F. acknowledges support from
the NSF under grant No. DMR-9973225. We also acknowledge support from
Georgetown University for a travel grant in the fall of 2001.

\end{document}